%% file: alicepreprint_CDS.tex
\documentclass[ALICE,manyauthors]{cernphprep}

\usepackage[comma,square,numbers,sort&compress]{natbib}
\usepackage{lineno}
\usepackage{rotating}
\usepackage{color}
\usepackage{enumerate}
\usepackage{hyperref}
\hypersetup{
    colorlinks=true,
    linkcolor=red,          
    citecolor=red,        
    filecolor=magenta,      
    urlcolor=black           
}
\graphicspath{{./figs/}}

\begin{document}%

\begin{titlepage}
\PHyear{2015}
\PHnumber{082}      
\PHdate{28 March}  
%

\title{Coherent $\rho^0$ photoproduction in ultra-peripheral Pb--Pb collisions \\ 
at $\mathbf{\sqrt{\textit{s}_{\rm NN}}=2.76}$ TeV}
\ShortTitle{Coherent $\rho^0$ photoproduction}   

\Collaboration{ALICE Collaboration\thanks{See Appendix~\ref{app:collab} for the list of collaboration members}}
\ShortAuthor{ALICE Collaboration} 

\begin{abstract}
We report the first measurement at the LHC of coherent photoproduction of $\rho^0$ mesons 
in ultra-peripheral Pb--Pb collisions. The invariant mass and transverse momentum 
distributions for $\rho^0$ production are studied in the $\pi^+ \pi^-$ 
decay channel at mid-rapidity. The production cross section in the rapidity 
range $|y| <$~0.5 is found to be 
$\mathrm{d}\sigma/\mathrm{d}y = 425 \pm 10 \, (\mathrm{stat.})$ $^{+42}_{-50} \, (\mathrm{sys.})$~mb. 
Coherent $\rho^0$ production is studied with and without requirement of nuclear breakup, 
and the fractional yields for various breakup scenarios are presented. 
The results are compared with those from lower energies and with model predictions. 
\end{abstract}
\end{titlepage}
\setcounter{page}{2}

\input{upcrho0.tex}               
%
%

\newenvironment{acknowledgement}{\relax}{\relax}
\begin{acknowledgement}
\section*{Acknowledgements}
\input{acknowledgements_jan2015.tex}    
\end{acknowledgement}

\bibliographystyle{utphys}   
\bibliography{upcrho0}

\newpage
\appendix
\section{The ALICE Collaboration}
\label{app:collab}
\input{Alice_Authorlist_2015-Mar-13_mod.tex}  
\end{document}

%% file: upcrho0.tex
\section{Introduction}

Charged particle beams at the LHC generate an electromagnetic field which 
can be regarded as a beam  of quasi--real photons; thus at the LHC, besides 
hadronic interactions, also photonuclear and photon--photon interactions occur. 
Collisions in which the impact parameter exceeds the sum of the radii of the 
incoming beam particles are called ultra--peripheral collisions (UPC).
In UPC the cross section for hadronic processes is strongly suppressed, while the cross 
sections for two--photon and photonuclear interactions remain large.
This is particularly the case for heavy ions, because the intensity of the photon flux grows 
with the square of the ion charge, $Z$. A number of reviews of 
UPC are available; e.g., \cite{Baltz:2007kq,Bertulani:2005ru}. The ALICE Collaboration has 
previously studied exclusive photoproduction of $J/\psi$ in ultra-peripheral Pb--Pb and p--Pb 
collisions~\cite{Abelev:2012ba,Abbas:2013oua,TheALICE:2014dwa}. 

Exclusive photoproduction of $\rho^0$ vector mesons, $\mathrm{Pb+Pb} \rightarrow \mathrm{Pb+Pb}+\rho^0$, can 
be described as the fluctuation of a quasi-real photon into a virtual $\rho^0$ vector meson, 
which then scatters elastically off the target nucleus. Two cases can be 
distinguished. When the interaction involves the complete target nucleus, 
the process is called coherent. In this case the target nucleus normally 
remains intact. If the virtual  $\rho^0$ vector meson scatters off only 
one of the nucleons in the target, then the
process is called incoherent and in this case the target nucleus normally 
breaks up, emitting neutrons at very forward rapidities. For coherent 
processes, the size of the lead ion restricts the mean transverse momentum of the vector 
meson to be about 60~MeV/$c$ corresponding to a de Broglie wavelength 
of the nuclear size, while it is of the order of 500 MeV/$c$ for incoherent processes.

Because of the strong electromagnetic fields in ultra-peripheral collisions of 
heavy ions, multiple photons may be exchanged in a single event. The additional 
photons can lead to excitation of the nuclei. The dominant process is 
the excitation to a Giant Dipole Resonance~\cite{Berman:1975tt}. As these 
photonuclear processes occur on a different time scale, they are assumed to 
be independent, so the probabilities factorize. The excited nucleus typically 
decays by the emission of neutrons at 
very forward rapidities. The signature of these processes is thus a 
$\rho^0$ vector meson with very low transverse momentum which may be accompanied 
by a few neutrons at very forward rapidities but no other particles. 

Photoproduction of $\rho^0$ vector mesons on nuclear targets has been studied in 
fixed target experiments with lepton beams~\cite{Mcclellan:1972cz}, 
and more recently in ultra-peripheral collisions by the STAR Collaboration at RHIC at 
$\sqrt{s_{\rm NN}} =$ 62~\cite{Agakishiev:2011me}, 130~\cite{Adler:2002sc}, 
and $200$ GeV~\cite{Abelev:2007nb}. STAR has also observed 
coherent photoproduction of the $\rho^0(1700)$  \cite{Abelev:2009aa}.

The $\rho^0$ vector meson gives the dominant contribution to the hadronic structure of 
the photon. For proton targets, the process $\gamma + p \rightarrow \rho^0 +p$ contributes 
about 10\% to 20\% of the total $\gamma+p$ cross section, depending on energy~\cite{Crittenden}. 
Scaling from a nucleon target to a nuclear target is often done using the Glauber model 
assuming Vector Meson Dominance~\cite{Bauer:1977iq}. 
The large value of $\sigma(\gamma + \mathrm{p} \rightarrow \rho^0 +\mathrm{p})$ means that for heavy nuclei 
one may reach the limit where the target appears like a black disk and 
the total $\rho^0+A$ cross section approaches $2 \pi R_A^2$ ($R_A$ is the nuclear radius). 
The situation may, however, be more complicated for several reasons. 
The cross section $\sigma(\gamma + \mathrm{p} \rightarrow \rho^0 +\mathrm{p})$ has contributions both 
from Reggeon and Pomeron exchange, and its energy dependence is therefore not monotonic. 
Furthermore, the nuclear medium might modify the Reggeon and Pomeron components differently.  
There may also be interference between the $\rho$ and $\rho'$ production amplitudes, 
and these amplitudes may be affected by the nuclear environment in a different way~\cite{Pautz:1997eh}. 
A detailed discussion of models for photoproduction of $\rho^0$ on complex nuclei based 
on data from fixed target experiments can be found in~\cite{Bauer:1977iq}. 

The cross sections measured by STAR~\cite{Adler:2002sc,Abelev:2007nb,Agakishiev:2011me} at RHIC were 
found to be about a factor two less than that predicted by the calculation 
of Ref.~\cite{Frankfurt:2002wc}, while in agreement with STARLIGHT~\cite{Klein:1999qj}. 
The reason for the difference between these two models, which both 
use the Glauber model to obtain the $\gamma$-nucleus cross section, will be discussed below. 
The many issues associated with calculating the photonuclear $\rho^0$ cross section and the 
discrepancies between models thus call for more data. In particular, it is important to establish if 
the trends seen at lower energies persist at higher energies. 

Moreover, the total cross section for exclusive $\rho^0$ production is very large at LHC energies, with the 
models mentioned above predicting that it could be between 50--100\% of the total hadronic inelastic 
cross section. It could thus constitute a significant background, e.g. at the trigger level, to low 
multiplicity peripheral hadronic interactions and to other types of ultra-peripheral collisions. It 
therefore has to be well understood. The high statistics in the $\rho^0$ sample allows the predictions 
for exclusive $\rho^0$ production accompanied by nuclear fragmentation to be tested with good precision. 

This paper presents the first measurement of the cross section for coherent
photoproduction of $\rho^0$ vector mesons in Pb--Pb collisions at the LHC.
The $\rho^0$ is reconstructed using the $\pi^+\pi^-$ decay channel in the rapidity range $|y| <$~0.5. 
The rapidity interval corresponds to a $\gamma$-nucleon center of mass energy in the 
range $36 \leq W_{\gamma N} \leq 59$~GeV with $\langle W_{\gamma N} \rangle =$~48~GeV, about a factor of 4 higher 
than in any previous measurement~\cite{Abelev:2007nb}. The cross section is measured for the 
cases of no neutron emission and for at least one emitted neutron. The new data presented in
this paper will hopefully help to clarify some of the theoretical uncertainties mentioned above.

\section{The ALICE experiment and the UPC trigger}

A full description of the ALICE detectors and their performance can be found in
\cite{Aamodt:2008zz,Abelev:2014ffa}; here, only the components relevant for this analysis will 
be briefly described. The Inner Tracking System (ITS) and Time Projection Chamber 
(TPC) are used to measure and identify the tracks of the decay products of the $\rho^0$ vector 
meson. The ITS consists of six layers of silicon detectors covering the full azimuthal 
angle. The two innermost layers form the Silicon Pixel Detector (SPD)
with a pseudorapidity acceptance of $|\eta|<1.4$. 
The SPD also provides trigger information at the lowest level. 
Two layers of silicon drift and two of silicon strip detectors complement the ITS, and
all six layers have an acceptance of $|\eta|<0.9$. The TPC is 
the main tracking detector of ALICE. It has a Ne--CO$_2$--N$_2$ gas mixture contained in 
a large -- almost 90 m$^3$ -- cylindrical drift detector with a central membrane at 
high voltage and two readout planes, composed of multi-wire proportional chambers, at the 
end caps. It covers the 
full azimuth and $|\eta|<0.9$ for full length tracks. It also provides a measurement 
of the ionization energy loss, d$E/$d$x$, which allows the identification of particles.
The TPC and ITS are situated inside a large solenoid magnet providing a B = 0.5~T field. 

\begin{figure}[htbp]
\centering
{\includegraphics[width=0.6\linewidth]{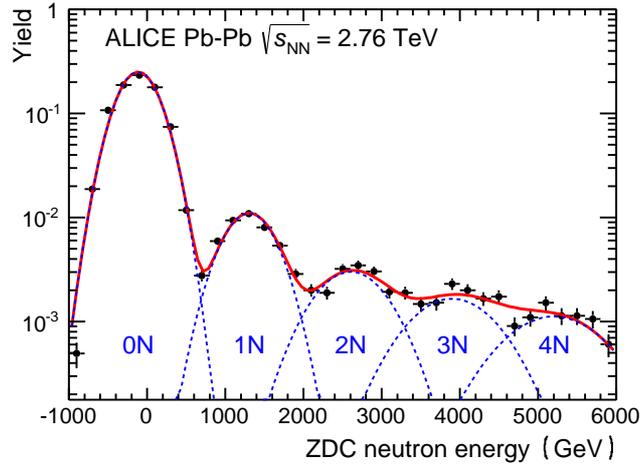}}
\caption{Energy deposit in the Zero-Degree Calorimeters. The curves correspond to Gaussian fits 
for 0, 1, 2, 3, or 4 neutrons entering the calorimeter. The plot is for events satisfying the 
requirements (i)--(vi) described in the text. 
}
\label{fig:zdc}
\end{figure}

The measurement of neutrons emitted at forward rapidities is performed with a set of two
neutron Zero Degree Calorimeters (ZDC) located 114 m away on each side of the
interaction point. The ZDC has a 99\% detection probability for neutrons
with $|\eta|>8.8$\cite{ALICE:2012aa}.
Figure~\ref{fig:zdc} illustrates the capabilities of the ZDC to
separate the emission of zero, one or several neutrons at zero
degrees. The sample appearing in this figure was obtained from events
fulfilling the event selection described in Section 3. 

In addition to the SPD, this analysis uses the Time of Flight (TOF) and  
VZERO detectors for triggering. TOF is a large cylindrical barrel surrounding the TPC. It has 18 sectors 
in azimuth, each made of
multigap resistive plate chambers distributed in five gas-tight modules, 
with a total of 152928 read-out channels and an intrinsic time resolution better than 50 ps.
The pseudorapidity acceptance is the same as for the TPC. The VZERO consists of two arrays of
scintillators called VZERO-A and VZERO-C, covering 
the pseudorapidity ranges $2.8<\eta <5.1$ (VZERO-A) and $-3.7<\eta <-1.7$ (VZERO-C). Its time 
resolution, better than 500~ps~\cite{Abbas:2013taa}, allows beam-beam collisions to be distinguished 
from beam-gas collisions.

The data used for this analysis were collected during the 2010 Pb--Pb
run of the LHC at an energy of $\sqrt{s_{\rm NN}} = 2.76$~TeV. Two different triggers
were used. At the beginning of the run, when the luminosity was
low, the trigger requirement was at least two hits in the TOF detector.
When the luminosity was increased the trigger selection was strengthened to 
improve the purity 
by additionally requiring at least two hits in the outer layer of the SPD, 
and no activity in any of the VZERO arrays. 

The luminosity is determined from the cross section for triggering on at least
one neutron in the ZDC detectors~\cite{ALICE:2012aa}. This cross section has been determined from a 
van der Meer scan~\cite{vanderMeer:1968zz} to 
be \linebreak $371.4 \pm 0.6 \, (\mathrm{stat.}) ^{+24}_{-19} \, (\mathrm{syst.})$~b~\cite{Abelev:2014ffa}.
The integrated luminosities for the two samples are $48 ^{+3}_{-2}$~mb$^{-1}$ (TOF trigger only) and 
$214 ^{+14}_{-11}$~mb$^{-1}$ (SPD+TOF+VZERO trigger).

\section{Track and event selection}

\begin{figure}[htbp]
\centering
{\includegraphics[width=0.55\linewidth]{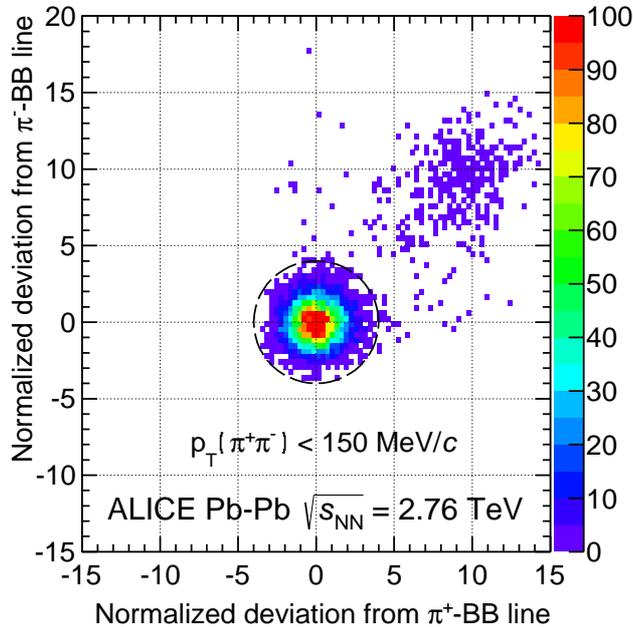}}
\caption{Identification of $\pi^{+/-}$ from the TPC d$E$/d$x$. The $x$-- and $y$--axes show the 
deviation of the measured energy loss from the Bethe-Bloch (BB) expectations for positive and 
negative tracks, respectively. The scale is such that one unit corresponds to one standard deviation. 
The circle, corresponding to $4 \sigma$, shows the selection used. The entries in the upper 
right corner are from $e^+e^-$ pairs produced in two-photon interactions. The plot is for events 
satisfying the requirements (i)--(iii) and (v) -- (vi) described in the text.  
}
\label{fig:dedx}
\end{figure}

In addition to the trigger selection, the events used for the analysis are required to fulfill 
the following requirements: 
\begin{enumerate}[i)]
\setlength{\itemsep}{0cm}
\setlength{\parskip}{0cm}
\item a primary vertex has to be identified within 10 cm of the nominal interaction point position, 
  along the beam direction; 
\item the event is required to have exactly two tracks reconstructed in the TPC and ITS satisfying 
  the track selections discussed below; 
\item the VZERO arrays are required to be empty (the difference between the offline and online 
VZERO selection will be discussed below); 
\item the energy loss in the TPC has to be consistent with that for pions within
  $4$ standard deviations from the Bethe-Bloch expectations, {\it i.e.}, 
  $\Delta\sigma_{\pi^+}^2+ \Delta\sigma_{\pi^-}^2 < 16$ (see Fig.~\ref{fig:dedx}); 
\item the track pairs used to define the coherent signal have to have a transverse momentum 
below $150$~MeV/$c$ and rapidity $|y| <$~0.5, the latter
requirement being imposed to avoid edge effects; 
\item the track pairs used to define the coherent signal are required to have tracks of opposite charge. 
\end{enumerate}

The background estimated from like-sign pairs ($\pi^+ \pi^+$ and $\pi^- \pi^-$) is below 2\% and it is 
subtracted from the final sample bin-by-bin in invariant mass. 

The track selection requires that each track has at least 70 space
points, out of a maximum of 159, in the TPC and a $\chi^2$  per degree of 
freedom from the Kalman fit procedure better than 4. 
Each track has at least one hit in the SPD with a $\chi^2$
per ITS hit less than 36. The distance of closest approach between the
track and the primary vertex has to be less than 2 cm along the beam
direction and less than $0.0182+0.035/p^{1.01}_T$ cm ($p_{\rm T}$ in GeV/c) 
in the plane perpendicular to the beam direction. These track selection cuts are 
based on studies of the detector performance~\cite{Abelev:2014ffa}. 

Three other track selections are used in order to estimate systematic
errors. These differ from the default track selection described
above in the following ways: (a) accepting tracks reconstructed only
in the ITS in addition to combined ITS-TPC tracks satisfying the default track selection;
(b) using only TPC information and accepting tracks having at least 50 space points
in the TPC; (c) using the default track cuts with
stronger requirements on TPC variables. The latter requirements meant that 
the tracks had to pass at least 120 of the 159 TPC pad rows and have a 
cluster in more than 80\% of the crossed pad rows. 
For the cross section calculation, 
the mean of the results of the four different track selection methods is used.  
The systematic error related to the track selection is estimated from the 
deviation from the mean. This contributes $^{+3.7}_{-3.0}$\% to the systematic
error.  

The momentum resolution of the ALICE central barrel tracking system~\cite{Abelev:2014ffa} 
translates into a resolution in transverse momentum of single $\pi^+\pi^-$--pairs better 
than 4 MeV/c in the kinematic range studied here. Similarly, the resolution in invariant 
mass varies between 2 MeV/$c^2$ ($M_{\pi \pi} = 0.4$~GeV/$c^2$) and 
6 MeV/$c^2$ ($M_{\pi \pi} = 1.5$~GeV/$c^2$). 

The ionization energy loss for the selected tracks is shown in 
Fig.~\ref{fig:dedx}. The scale on both axes is in units of 
the number of standard deviations from the Bethe-Bloch expectation in the TPC; in
this way the dependence on track momentum is removed. Pions can be
clearly identified by the $4\sigma$ circle centered on $(0,0)$, while the 
events above and to the right of the pions are mostly $e^+e^-$ pairs
from $\gamma\gamma$~$\rightarrow$~$e^+e^-$. This figure shows that any
possible contamination from kaons or protons in the sample is negligible. 
There could, however, be a contamination from muons from the 
process $\gamma \gamma \rightarrow \mu^+ \mu^-$, which cannot 
be distinguished from pions using the energy loss. This contribution can be 
estimated from the number of $e^+e^-$ pairs in the data sample, as the cross 
sections for $\gamma \gamma \rightarrow \mu^+ \mu^-$ and 
$\gamma \gamma \rightarrow e^+ e^-$ are about the same at midrapidity for 
invariant masses well above threshold. It can also be calculated  
from STARLIGHT~\cite{Baltz:2009jk,starlight}. Both methods give an expected number of
muon pairs of about 5\%, which is not corrected for, but added to the systematic error. 
The contribution from $\gamma \gamma \rightarrow \pi^+ \pi^-$ is expected to be much
smaller than from $\gamma \gamma \rightarrow \mu^+ \mu^-$. The $\pi^+ \pi^-$ cross section is
reduced by the form factor of the pion, see e.g. \cite{Klusek-Gawenda:2013rtu}, so this 
contribution is not considered.

\section{Data analysis}

\begin{figure}[htbp]
\begin{center}
\includegraphics[width=0.5\linewidth,keepaspectratio]{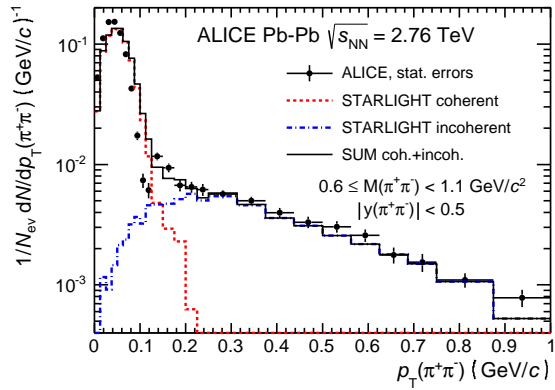}
\end{center}
\caption{Transverse momentum distributions for $\pi^+ \pi^-$--pairs.
The dashed (red) and dash-dotted (blue) histograms show the normalized $p_{\rm T}$ 
distribution from STARLIGHT passed through the detector response simulation for coherent 
and incoherent $\rho^0$ production, respectively. The solid (black) histogram is 
the sum of the two.}
\label{fig:pT}
\end{figure}

Using the event and track selection described in the previous section, the 
four--momenta of the two tracks are
constructed and pair variables are extracted. 
The resulting distribution of the pair transverse momentum is shown in Fig.~\ref{fig:pT} for 
events with $0.4$~$\leq$~$M_{\pi \pi}$~$\leq$~$1.1$~GeV/$c^2$ and $|y| <$~0.5. 
A peak at low transverse momentum ($p_{\rm T} < 0.15$~GeV/$c$), corresponding to coherent
production, is clearly seen. The distribution is compared with the
corresponding distributions from STARLIGHT~\cite{Klein:1999qj,starlight} events for coherent 
and incoherent $\rho^0$ production, processed through the
detector response simulation based on GEANT~3. The coherent peak is shifted to slightly lower $p_{\rm T}$ in 
data than that predicted by STARLIGHT.  A similar trend has been observed by
STAR at lower energies~\cite{Debbe:2012aa}. The shape of the coherent peak in the $p_{\rm T}$ distribution 
is determined by the nuclear form factor. The form factor used in STARLIGHT is consistent with what is 
obtained from elastic electron-nucleus scattering, which probes the charge content of the nucleus. 
Since the $\rho^0$ couples to both neutrons and protons, a possible explanation of this difference 
could thus be the presence of a ``neutron skin''. The effect, however, appears larger than what 
the current limit on the difference between neutron and proton radius in $^{208}\mathrm{Pb}$ (0.3 fm) 
allows~\cite{Abrahamyan:2012gp}, and is thus not fully understood. 
Data also show a dip around $p_T = 0.12$~GeV/$c$, which is not present in the model. The absence of this
dip in the model can be understood from the fact that in STARLIGHT the transverse momentum of the photon
is considered, and this reduces the dip one would expect from the form factor of the target nucleus alone. 
In a Glauber calculation, the transverse momentum distribution is determined from a Fourier transform 
of the nuclear profile function, see e.g. \cite{Bertulani:2005ru}, and the direct dependence on the form 
factor is only an approximation; this could also contribute to explaining the difference between 
STARLIGHT and data. The high-$p_{\rm T}$ tail of the distribution is very well described by the 
incoherent $p_{\rm T}$ spectrum from STARLIGHT. 

The transverse momentum distribution for coherent production may also be parameterized as an 
exponential, $\mathrm{d}N/\mathrm{d}t \propto \mathrm{exp}(bt)$ where $t = - p_T^2$. Fitting the ALICE
data to such a function 
gives $b = 428 \pm 6 (\mathrm{stat.}) \pm 15 (\mathrm{syst.})$~GeV$^2$/$c^2$. The systematic error 
has been obtained as the difference in slope between STARLIGHT events and STARLIGHT events processed
through the full detector simulation. The ALICE result can 
be compared with the corresponding measurement by STAR, where $b = 388 \pm 24$~GeV$^2$/$c^2$ was 
found~\cite{Abelev:2007nb}. The STAR and ALICE results are consistent within errors if one takes 
into consideration that $b$ is expected to be $\approx$4--8\% larger for a lead nucleus than for a gold
nucleus because of the difference in size (one expects $b \propto R^2$). The fit was performed for 
$|t| > 0.002$~GeV$^2$/$c^2$ to avoid interference effects at very low $p_T$~\cite{Abelev:2007nb}.

The final sample of coherent $\rho^0\to\pi^+\pi^-$ candidates is
corrected for acceptance and efficiency in invariant mass bins. The
event sample used to determine the corrections has uniform distributions in
invariant mass, rapidity, transverse momentum, and azimuthal angle
over the ranges $2m_{\pi}$~$\leq$~$M_{\pi \pi}$~$\leq$~$1.5$~GeV/$c^2$,
$|y|$~$\leq$~$1.0$, $p_{\rm T}$~$\leq$~$0.15$~GeV/$c$, and
$0$~$\leq$~$\phi$~$\leq$~$2\pi$.  Using a flat distribution in transverse 
momentum is justified over the narrow range $p_{\rm T}$~$\leq$~$0.15$~GeV/$c$, where the 
acceptance and efficiency are constant. All models predict only a very small variation 
of the cross section over the range $|y| <$~0.5 (see Fig.~\ref{fig:dsigdy} below) so also 
for rapidity a uniform input distribution is justified. The advantage of using a flat input 
distribution in invariant mass is to obtain sufficient statistics in the tails 
of the distribution. If one were to use a $\rho^0$--shape as input, one would 
need enormous statistics to cover the high and low invariant mass ranges. 
The $\rho^0$ candidates are assumed to be transversely polarized. This is expected from 
helicity conservation and has been confirmed by photoproduction 
measurements~\cite{Abelev:2007nb,Breitweg:1997ed}. This polarization translates into a 
$\mathrm{d}n_{\pi}/\mathrm{d}\Omega \propto \sin^2(\theta)$ angular distribution of the $\pi^+ \pi^-$ decay 
products in their center of mass system ($\theta$ is here measured relative to the direction of 
flight of the $\rho^0$ in the $\gamma$-nucleon center of mass system). 
All generated samples serve as input to a full detector simulation using GEANT~3 for the 
propagation of particles through the detector. Selection criteria are applied in the same way 
as done for real events. The variation of the detector configuration during the 
data taking period is included in the detector response simulations.  
The product of acceptance and efficiency varies from about 2\% at the low end of 
the studied invariant mass interval ($M_{\pi \pi} =$~0.6~GeV/$c^2$) to 
about 12\% at the high end ($M_{\pi \pi} =$~1.5~GeV/$c^2$).

\begin{figure}[htbp]
\begin{center}
{\includegraphics[width=0.5\linewidth,keepaspectratio]{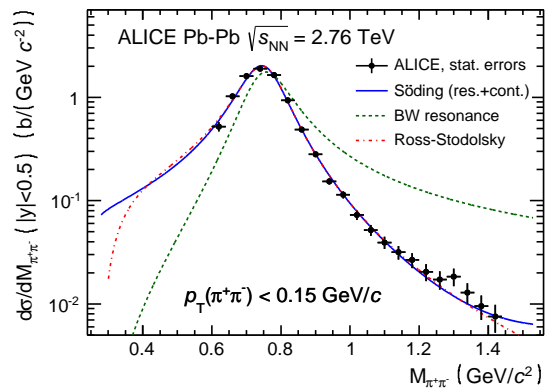}}
\end{center}
\caption{Invariant mass distribution for pions corrected for acceptance and efficiency. 
The solid (blue) curve corresponds to a fit to the S\"oding parameterization Eq.~\ref{eq:minv}, 
and the dashed (green) curve shows the resonant contribution only. The dot-dashed (red) curve shows 
the fit to the Ross-Stodolsky parameterization Eq.~\ref{minv:rs}. 
The parameters of the fit are given in the text. } 
\label{fig:invmass_pions}
\end{figure}

The uncertainty in the trigger efficiency is obtained by comparing the measured trigger 
efficiency with the one in the detector response simulation in a data sample taken with 
a ZDC trigger~\cite{Abbas:2013oua}. The result is a trigger efficiency uncertainty 
of $^{+3.8}_{-9.0}$\%. In addition, a correction 
is applied for the trigger dead time resulting from after pulses in the TOF. The 
systematic error on this correction is estimated to be $\pm 1.3$\%. 

The invariant mass distribution of the $\rho^0$ candidates, corrected for acceptance 
and efficiency and normalized by the luminosity to provide a cross section, is shown
in Fig.~\ref{fig:invmass_pions}. It is well known that the shape of the $\rho^0$ 
in photoproduction deviates from a pure Breit-Wigner 
resonance~\cite{Mcclellan:1972cz,Adler:2002sc,Abelev:2007nb,Agakishiev:2011me,Breitweg:1997ed}. 
Several different parameterizations 
exist to describe the shape, with one of the most often used being a formula due to S\"oding, 
where a continuum amplitude, $B$, is added to the Breit-Wigner resonance~\cite{Soding:1965nh}:
\begin{equation}
\frac{\mathrm{d}\sigma}{\mathrm{d}M_{\pi\pi}} =
\left|
A\frac{\sqrt{M_{\pi\pi} M_{\rho^0} \Gamma(M_{\pi\pi})}}
{M^2_{\pi\pi} - M^2_{\rho^0} + i M_{\rho^0} \Gamma(M_{\pi\pi})}+B
\right|^2 \; .
\label{eq:minv}
\end{equation}
Here, $A$ is the amplitude of the Breit--Wigner function, $B$ is the 
amplitude of the direct non--resonant $\pi^+\pi^-$ production, and the mass dependent 
width is given by 
\begin{equation}
\Gamma(M_{\pi\pi}) = \Gamma_{\rho^0} \frac{M_{\rho^0}}{M_{\pi\pi}}
\left[
\frac{M^2_{\pi\pi}-4m^2_\pi}{M^2_{\rho^0}-4m^2_\pi}
\right]^{\frac{3}{2}}
\; ,
\end{equation}
with $m_\pi$ the mass of the pion.  Eq.~\ref{eq:minv} was fitted to the measured 
$M_{\pi\pi}$ distribution with $M_{\rho^0}$, $\Gamma_{\rho^0}$, $A$, and 
$B$ as free parameters. The fit gives 
$M_{\rho^0} = 761.6 \pm 2.3 \, (\mathrm{stat.}) ^{+6.1}_{-3.0} \, (\mathrm{syst.})$~MeV/$c^2$ and 
$\Gamma_{\rho^0} = 150.2 \pm 5.5 \, (\mathrm{stat.}) ^{+12.0}_{-5.0} \, (\mathrm{syst.})$~MeV/$c^2$, 
in agreement with the values reported by the PDG~\cite{Beringer:1900zz}. 
The ratio of the non--resonant and resonant amplitudes is found to be 
$|B/A| = 0.50 \pm 0.04 \, (\mathrm{stat.}) ^{+0.10}_{-0.04} \, (\mathrm{syst.})$ (GeV/c$^2$)$^{-1/2}$. 
The systematic errors are obtained by varying the fitting method 
($\chi^2$ or log likelihood minimization), track selection (as discussed above), 
and fitting range. 

The ratio $|B/A|$ is lower than what was found by STAR with Au targets, 
$|B/A| =$~0.81--0.89~(GeV/c$^2$)$^{-1/2}$ for $\langle W_{\gamma N} \rangle$ in the range 
7--12~GeV~\cite{Adler:2002sc,Abelev:2007nb,Agakishiev:2011me}. The result 
from ZEUS with proton targets for $\langle W_{\gamma N} \rangle$ in the range 55--90~GeV shows that $|B/A|$ 
varies with the $\gamma$--proton momentum tranfer~\cite{Breitweg:1997ed}. The average 
is $|B/A| = 0.67 \pm 0.02 \, (\mathrm{stat.}) \pm 0.04 \, (\mathrm{syst.})$, 
while for momentum transfers of the same order as for coherent 
production $|B/A| \approx$~0.8. The lower value of $|B/A|$ 
observed by ALICE may indicate that the non-resonant contribution is more strongly absorbed 
in heavy nuclei at high energies, as had been previously suggested~\cite{Bauer:1971sz}. 

Other parameterizations of the $\rho^0$ shape are possible, 
and as a cross check 
the invariant mass distribution was also fit to a Ross-Stodolsky 
function~\cite{Breitweg:1997ed,Ross:1965qa}: 
\begin{equation}
\frac{\mathrm{d}\sigma}{\mathrm{d}M_{\pi\pi}} =
f \; 
\left|
\frac{\sqrt{M_{\pi\pi} M_{\rho^0} \Gamma(M_{\pi\pi})}}
{M^2_{\pi\pi} - M^2_{\rho^0} + i M_{\rho^0} \Gamma(M_{\pi\pi})}
\right|^2 
\left( \frac{M_{\rho^0}}{M_{\pi\pi}} \right)^k \; \; ,
\label{minv:rs}
\end{equation}
with a slightly different definition of the mass dependent width
\begin{equation}
\Gamma(M_{\pi\pi}) = \Gamma_{\rho^0} 
\left[
\frac{M^2_{\pi\pi}-4m^2_\pi}{M^2_{\rho^0}-4m^2_\pi}
\right]^{\frac{3}{2}} \; \; .
\end{equation}
As can be seen in Fig.~\ref{fig:invmass_pions}, this parameterization also 
described the observed shape of the invariant mass distribution well and gave a $\rho^0$ 
mass ($M_{\rho^0} = 769.2 \pm 2.8 \, (\mathrm{stat.}) ^{+8.0}_{-5.2} \, (\mathrm{syst.})$~MeV/$c^2$) and width 
($\Gamma_{\rho^0} = 156.9 \pm 6.1 \, (\mathrm{stat.}) ^{+17.3}_{-5.9} \, (\mathrm{syst.})$~MeV/$c^2$) consistent with 
the PDG values. The deviation from a pure Breit-Wigner shape is given by the parameter 
$k$, which was found to be $k = 4.7 \pm 0.2 \, (\mathrm{stat.}) ^{+0.8}_{-0.6} \, (\mathrm{syst.})$. 
This can be compared to the corresponding value for proton targets from ZEUS~\cite{Breitweg:1997ed} and 
H1~\cite{Aid:1996bs} at HERA. ZEUS finds $k=5.13 \pm 0.13$ averaged over all momentum transfers and $k \approx 6$ for 
$t=0$, while H1 reports $k = 6.84 \pm 1.00$ averaged over all momentum transfers. The larger value of $k$ for
proton targets again indicates that the invariant mass distribution for Pb-targets deviates less from a pure
Breit-Wigner resonance, as was also found using the S\"oding formula. 

As can be seen in the lower part of Fig.~\ref{fig:invmass_pions}, there is a hint of a 
resonance around 1.3 GeV/c$^2$. This may be understood from two-photon production of the $f_2(1270)$ 
meson followed by its decay into two pions, $\gamma+\gamma \rightarrow f_2(1270) \rightarrow \pi^+ \pi^-$.  
This meson is a ``standard candle'' in two-photon interactions with a well known $\gamma \gamma$ coupling, 
but it has so far not been 
observed in ultra-peripheral collisions because of the large background from photonuclear processes. 
The significance of the excess over the $\rho^0$ Breit-Wigner distribution is estimated to be $4^{+2}_{-1}$, 
where the error comes from the uncertainty in the skewness of the Breit-Wigner distribution (parameter $k$ 
in the Ross-Stodolsky formula). 

The normalized yield of $\rho^0$s ($N_{\mathrm{yield}}$) is obtained by integrating the resonant part of 
Eq.~\ref{eq:minv} (obtained by setting $B = 0$ and taking the other parameters from the fit) 
from $2m_{\pi}$ to 1.5 GeV/c$^2$. 
The systematic error on the number of extracted $\rho^0$s 
is obtained by varying the fitting method ($\chi^2$ or log likelihood minimization) 
and fitting range, resulting in an error of $^{+0.8}_{-1.4}$\%. 
The uncertainty in the track selection gives an additional error of $^{+3.7}_{-3.0}$\% as discussed above. 
Both Eq.~\ref{eq:minv} and \ref{minv:rs} describe the observed shape equally well (the integrated yield differ
by less than 0.5\%), so no additional systematic error was added to the yield because of the choice of fitting
function. 

It is worth noting that the shape of the resonant contribution (shown by the dashed curve in 
Fig.~\ref{fig:invmass_pions}) is quite different 
from the shape of the measured $\pi^+\pi^-$ invariant mass distribution. However, the integrated 
yield between $2m_{\pi}$ and 1.5 GeV/c$^2$ does not deviate by more than 
around 1\% if the non-resonant amplitude is included in the integration. 

The number of extracted $\rho^0$s is corrected for the following 3 contributions:  
incoherent events with $p_{\rm T} <$~0.15~GeV/c ($f_{\mathrm{incoh}}$), events which have one 
or more additional SPD tracklets ($f_{\mathrm{SPD}}$), and the number of coherent $\rho^0$ events 
lost by the VZERO offline timing requirement ($f_{\mathrm{VZERO}}$). 

The number of incoherent events with $p_{\rm T} <$~0.15 GeV/c is estimated in two 
different ways: first fitting the sum of two exponentials in $p_{\rm T}^2$ to the 
$p_{\rm T}$ distributions and integrating 
the fitted functions over the interval chosen for the coherent selection ($p_{\rm T} <$~0.15 GeV/c), and 
second using the fit to the STARLIGHT templates shown in Fig.~\ref{fig:pT}. 
The correction for incoherent events is found to be 5.1\% in both cases with an uncertainty estimated from 
using different track selections of $\pm$0.7\%. 

The track selection (a) above allows one to check the events for any additional 
activity in the ITS, for example from tracks with low momenta, which do not reach the TPC, 
using SPD tracklets, defined as any combination of hits from the two SPD layers. 
Rejecting events with one or more extra tracklets, not associated with the two good tracks 
coming from the primary vertex, removes 3.0\% of the events in the signal region. Since 
true UPC events should have no additional tracks, the extracted yield is corrected for this. 
In the Monte Carlo samples of coherently produced $\rho^0$s, the same cut removes only 
0.5\% of the events which is taken as the systematic error associated with this cut. 

The events selected by the SPD+TOF+VZERO trigger are required to have no online signal in the 
VZERO detector. A similar cut is also applied offline to the events triggered by TOF only. 
The VZERO offline selection is further refined using the timing information. This selection has 
been tuned to work well for hadronic interactions, which typically have 
a non-zero signal in the VZERO on both sides. In the ultra-peripheral 
events studied here, where the VZERO is required to be empty, the offline selection is less reliable, 
and a coherent signal can be observed in the events with 2 tracks rejected by the offline VZERO requirement. 
The increase in the coherent signal when the offline VZERO selection is not used 
amounts to 10.0\%. The systematic error of this number is obtained from the estimated contamination 
from hadronic events following from this looser cut. This contamination is determined from the fraction 
of the events which have a signal in the ZDCs, resulting in a systematic error of $^{+0.0}_{-3.1}$\%.

The corrected number of coherent $\rho^0$s is then obtained from 
\begin{equation}
N_{\rho}^{coh} = \frac{N_{\mathrm{yield}}}{1 + f_{\mathrm{incoh}} + f_{\mathrm{SPD}} + f_{\mathrm{VZERO}} } \; ,
\end{equation}
with $f_{\mathrm{incoh}} = 0.051 \pm 0.007$, $f_{\mathrm{SPD}} = 0.030 \pm 0.005$, and 
$f_{\mathrm{VZERO}} = -0.100 ^{+0.031}_{-0.000}$. 
From this number the differential cross section is calculated as 
\begin{equation} 
\frac{\mathrm{d}\sigma}{\mathrm{d} y}   =
\frac{N_{\rho}^{coh}}{\mathrm{L}_{int} \cdot \Delta y} \; .
\label{eq:xsection}
\end{equation} 

The systematic errors discussed above are summarized in Table~\ref{tab:systerror}. They have been 
evaluated for the SPD+TOF+VZERO trigger sample, which contains more than 80\% of the total integrated 
luminosity. The total error is obtained by adding the individual errors following the description 
in~\cite{Barlow:2003sg}. The two trigger samples, with appropriate errors, are compared as a cross check. 
They use different trigger combinations and 
were taken under quite different running conditions, with the typical hadronic minimum bias interaction 
rate being around 10~Hz during the early part of the run when the TOF only trigger was used and 
around 200~Hz during the later part of the run when the SPD+TOF+VZERO trigger was used. The correction 
factor for trigger dead time due to after pulses was thus very different for the two samples 
($\approx$1 during the early part and $\approx$5 during the later part). 

\begin{table}
\setlength{\extrarowheight}{5pt}
\begin{center}
\begin{tabular}{lc}
Variable                       & Systematic error   \\ \hline 
Luminosity                     & $^{+6.5\%}_{-5.1\%}$  \\
Trigger efficiency             & $^{+3.8\%}_{-9.0\%}$  \\ 
Trigger dead time correction   & $\pm 1.3$\%        \\ 
Signal extraction              & $^{+0.8\%}_{-1.4\%}$  \\
Track selection                & $^{+3.7\%}_{-3.0\%}$  \\
Particle ID                    & $^{+0.0\%}_{-5.0\%}$  \\ 
Incoherent contribution        & $\pm 0.7$\%        \\
SPD tracklets                  & $\pm 0.5$\%        \\
VZERO offline selection        & $^{+0.0\%}_{-3.1\%}$  \\ \hline 
Total                          & $^{+9.2\%}_{-11.2\%}$ \\ \hline 
\end{tabular}
\end{center}
\caption{\label{tab:systerror} 
Summary of the systematic error in the cross section calculation. The numbers are for the 
SPD+TOF+VZERO trigger sample. For a discussion of the TOF only trigger sample and the 
separation between correlated and uncorrelated errors of the two samples, see the text. 
}
\end{table}

To make a comparison of the cross sections measured under the different trigger conditions, 
the systematic errors are separated into correlated and uncorrelated errors for 
the two trigger samples. The fully correlated errors are those related to luminosity, incoherent contribution, 
trigger efficiency, and particle identification. The fully uncorrelated errors are those related to the VZERO 
offline selection (different VZERO thresholds were used for the two data samples), the cut on SPD tracklets, and 
trigger dead time. The errors related to the signal extraction and track selection are found to be partly 
correlated, but are decorrelated for the comparison. 
This gives a cross section 
$\mathrm{d}\sigma/\mathrm{d}y = 466 ^{+25}_{-25}$~mb for the sample taken with the TOF only trigger and 
$\mathrm{d}\sigma/\mathrm{d}y = 414 ^{+14}_{-16}$~mb for the sample taken with the SPD+TOF+VZERO trigger. 
The error is obtained from the squared sum of the statistical and uncorrelated systematic error. 
The difference of 12\% corresponds to 1.8 standard deviations. The final cross section is obtained as 
the weighted mean of the cross sections of the two samples. 
The weighting procedure provides a total error, including both the statistical and uncorrelated systematic 
components. The uncorrelated component is separated from the total error by subtracting in quadrature the
error obtained in the case when only the statistical errors are used for the weighting. 
The uncorrelated systematic error is then added in quadrature to the correlated 
systematic error to obtain the total systematic error. The final result is 
$\mathrm{d}\sigma/\mathrm{d}y = 425 \pm 10 \, (\mathrm{stat.}) ^{+42}_{-50} \, (\mathrm{syst.})$~mb. 

In addition to the $\rho^0$ cross section, the cross section for two-photon production of $e^+e^-$ pairs in the range 
$0.6 \leq M_{ee} \leq 2.0$~GeV/c$^2$  and $|\eta_{1,2}| <$~0.9 ($\eta_{1,2}$ are the pseudorapidities 
of the two tracks) was measured. The analysis is similar to the one for $\rho^0$ but the PID requirement 
was modified to accept electrons rather than pions. The detector efficiency is determined using 
STARLIGHT events processed through the full ALICE detector simulation. The result is 
$\sigma(0.6 \leq M_{ee} \leq 2.0 \; \mathrm{GeV}, |\eta_{1,2}| < 0.9) = 9.8 \pm 0.6 \, (\mathrm{stat.}) ^{+0.9}_{-1.2} \, (\mathrm{syst.})$~mb, 
which is in good agreement with the STARLIGHT~\cite{Baltz:2009jk} prediction for the same selection in 
invariant mass and pseudorapidity 
($\sigma =$~9.7~mb). The cross sections for the individual trigger samples are 
$11.8 \pm 1.6 \, (\mathrm{stat.}) ^{+1.1}_{-1.4} \, (\mathrm{syst.})$~mb (TOF only trigger) and 
$9.4 \pm 0.7 \, (\mathrm{stat.}) ^{+0.9}_{-1.1} \, (\mathrm{syst.})$~mb (SPD+TOF+VZERO trigger). 

As discussed above, photoproduction of vector mesons may occur in 
interactions where additional photons are exchanged between the nuclei, 
leading to neutron emission in the forward region. These neutrons may 
be detected in the ALICE ZDCs. Four Gaussian distributions centered 
around each peak with means and variances constrained to $x_n=nx_1$ 
and $\sigma_n=\sqrt{n}\sigma_1$ have been fitted to the ZDC energy 
distribution shown in Fig.~\ref{fig:zdc}. Here, $x_1$ and $\sigma_1$ 
are the position and width of the peak corresponding to one neutron, 
and $n$ is the number of neutrons. In order to separate 
different cases of neutron emission, the minima between the first 
three Gaussians are used.  The minimum between zero and one-neutron 
emission lies at half
the energy per nucleon and it is roughly three sigma away from the
adjacent peaks.  A given event is considered to have no neutron in
the ZDC if the energy registered in the calorimeter is less
than $600$~GeV, one neutron if the energy lies between $600$~GeV and
$2000$~GeV and more than one neutron if the energy is above $2000$~GeV.

The events are divided into different groups as follows: no neutrons 
emitted in any direction (0n0n), at least one neutron emitted in any 
direction (Xn), at least one neutron emitted in one direction and no 
neutron emitted in the other direction (0nXn), at least one neutron 
emitted in both directions (XnXn).  

The corrections applied in obtaining the cross section from the measured 
yield are independent of the ZDC signal.
The fractional yield for each fragmentation selection thus reflects the 
relative $\rho^0$ production cross section. The only exception to this is the 
correction for the incoherent contribution ($f_{\mathrm{incoh}}$), which is expected 
to be higher when a signal is required in the ZDCs. This correction is thus calculated 
for each ZDC selection separately, using the same method as described above. 

\begin{figure}
\begin{center}
\includegraphics[width=0.6\linewidth,keepaspectratio]{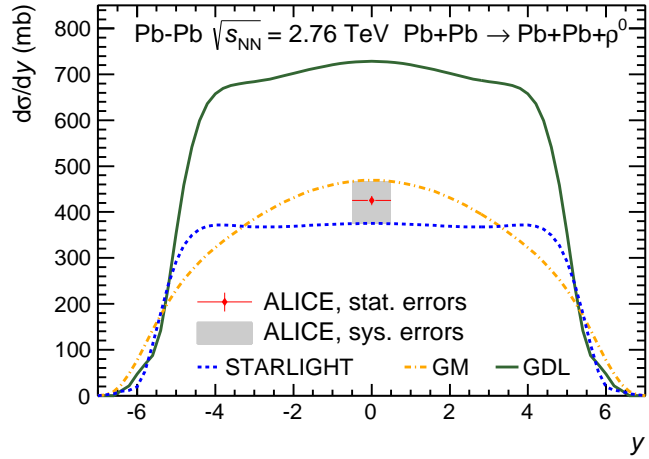}
\end{center}
\caption{The cross section for coherent photoproduction of $\rho^0$, $\mathrm{d} \sigma/\mathrm{d}y$, 
in ultra-peripheral collisions for the three models compared with the ALICE result. 
}
\label{fig:dsigdy}
\end{figure}

\section{Results and discussion}

The coherent $\rho^0$ photoproduction cross section, $\mathrm{d}\sigma/\mathrm{d}y$, is shown and compared 
with model predictions in Fig.~\ref{fig:dsigdy}. 
The measured cross section is in agreement with STARLIGHT~\cite{Klein:1999qj} and the calculation by Gon\c{c}alves 
and Machado (GM)~\cite{Goncalves:2011vf}, while the GDL (Glauber-Donnachie-Landshoff) 
prediction~\cite{Frankfurt:2002wc,Rebyakova:2011vf} is about a factor of 2 higher than data. The calculation by 
GM is based on the Color Dipole model, while STARLIGHT and GDL use the photon-proton cross section 
$\sigma(\gamma+\mathrm{p} \rightarrow \rho^0+\mathrm{p})$ constrained from data as input. 
In STARLIGHT, the $\gamma$-nucleon cross 
section is given by the parameterization $\sigma = 5.0 \, W_{\gamma N}^{0.20} + 26.0 \, W_{\gamma N}^{-1.23} \, \mu\mathrm{b}$ 
($W_{\gamma N}$ in GeV), while GDL use the Donnachie-Landshoff model~\cite{Donnachie:1999yb} for the total $\rho N$ 
cross section. 
All calculations use the Glauber model to scale the cross section from $\gamma$-nucleon to $\gamma$-nucleus. 

\begin{figure}
\begin{center}
\includegraphics[width=0.6\linewidth,keepaspectratio]{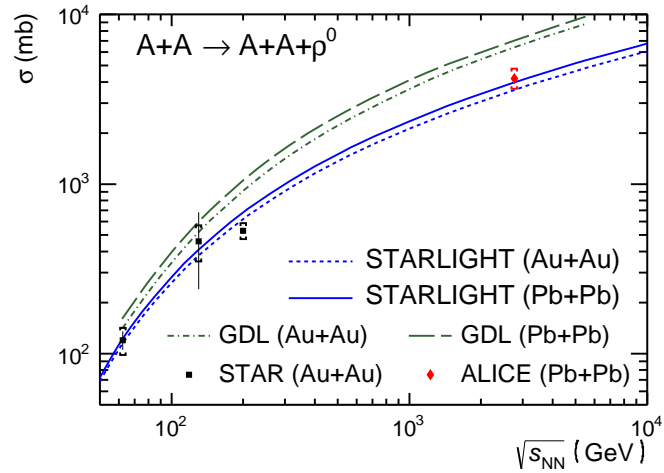}
\end{center}
\caption{Excitation function for coherent and exclusive $\rho^0$ production. The results from ALICE 
and STAR\cite{Adler:2002sc,Abelev:2007nb,Agakishiev:2011me} are compared with the STARLIGHT and GDL 
predictions for Pb--Pb and Au--Au.}
\label{fig:sig}
\end{figure}

The STAR Collaboration has published the total coherent $\rho^0$ photoproduction cross section at three different 
energies~\cite{Adler:2002sc,Abelev:2007nb,Agakishiev:2011me}. To be able to compare the current result to those, 
one has to integrate $\mathrm{d}\sigma/\mathrm{d}y$ 
over the whole phase space, which can only be done using models. The extrapolation factor from $|y|<0.5$ to
all rapidities is calculated as the mean of the values obtained from the STARLIGHT (10.6) and GM (9.1) models, 
and the deviation of the two from the mean ($\approx 8$\%) is added to the systematic error. 
This gives 
$\sigma(\mathrm{Pb+Pb} \rightarrow \mathrm{Pb+Pb}+\rho^0) = 4.2 \pm 0.1 (\mathrm{stat.}) ^{+0.5}_{-0.6} (\mathrm{syst.})$~b at 
$\sqrt{s_{\rm NN}} = 2.76$~TeV. The total cross section as a function of $\sqrt{s_{\rm NN}}$ is shown in 
Fig.~\ref{fig:sig}, where the results from ALICE and STAR Collaborations are compared with the STARLIGHT and 
GDL calculations. The total cross section increases by about a factor of 5 between the top RHIC energy 
and $\sqrt{s_{\rm NN}} = 2.76$~TeV.

The cross section and its energy dependence is well described by STARLIGHT, while the GDL calculation 
overpredicts the cross section by about a factor of 2. The agreement with STARLIGHT is somewhat 
surprising since its Glauber calculation 
does not include the elastic part of the total cross section, which is included in the GDL model. 
It has been argued that coherent $\rho^0$ production off heavy nuclei may probe the 
onset of the Black Body Limit, in which the total $\rho^0$--nucleus cross section approaches $2 \pi R_A^2$ at high 
energies~\cite{Frankfurt:2002wc}. The results from STAR and ALICE do not favour this picture. The cross section 
is instead reduced by about a factor of 2 compared with the GDL model~\cite{Rebyakova:2011vf}, independent of energy,
indicating that further work is needed to understand this process. 
It should be noted that none of the models in Fig.~\ref{fig:dsigdy} include cross terms such as 
$\rho + \mathrm{N} \rightarrow  \rho' + \mathrm{N}$. 

The photonuclear cross section, $\sigma(\gamma+\mathrm{Pb} \rightarrow \rho^0+\mathrm{Pb})$, in STARLIGHT is 
almost energy 
independent for $W_{\gamma N} > 10$~GeV. The increase in the Pb--Pb cross section, 
$\sigma(\mathrm{Pb+Pb} \rightarrow \mathrm{Pb+Pb}+\rho^0)$, with $\sqrt{s_{\rm NN}}$ is thus almost entirely due to 
the increase in the photon flux at higher collision energies. 

The model by GM, although in agreement with the current result, has been criticized for using the Color 
Dipole model for a soft probe like the $\rho^0$~\cite{Rebyakova:2011vf}. A recent publication shows that the 
calculation indeed has large uncertainties arising from the choice of $\rho^0$ wave function and dipole cross 
section~\cite{Santos:2014vwa}. 

The number of events satisfying the different fragmentation scenarios as well as the ratio to the total 
number of events are shown in Table~\ref{tab:zdc}. The table also shows the expected fractions from the 
STARLIGHT~\cite{Baltz:2002pp} and GDL~\cite{Rebyakova:2011vf} models.
These models assume that the probabilities for exchange of multiple 
photons in a single event factorize in impact parameter space. 
One should note that some of the fractions are 
correlated: the sum of (0n0n) and 
(Xn) should be 100\%, and the sum of (0nXn) and (XnXn) should be equal to 
(Xn). This is the case within errors, but the sum is not exact, since 
the incoherent contribution is subtracted for each selection separately. 
The results in Table~\ref{tab:zdc} are consistent with both the STARLIGHT and GDL 
calculations within three standard deviations.

\begin{table}
\setlength{\extrarowheight}{7pt}
\begin{center}
\begin{tabular}{lcccc}
Selection              & Number of events  & Fraction                                         & STARLIGHT & GDL    \\ \hline 
All events             & 7293              & 100 \%                                           &           &        \\ 
0n0n                   & 6175              & 84.7$\pm$0.4(stat.)$_{-1.9}^{+0.4}$(syst.) \%     & 79 \%     & 80 \%  \\ 
Xn                     & 1174              & 16.1$\pm$0.4(stat.)$_{-0.5}^{+2.2}$(syst.) \%     & 21 \%     & 20 \%  \\ 
0nXn                   & 958               & 13.1$\pm$0.4(stat.)$_{-0.3}^{+0.9}$(syst.) \%     & 16 \%     & 15 \%  \\ 
XnXn                   & 231               & 3.2$\pm$0.2(stat.)$_{-0.1}^{+0.4}$(syst.) \%      & 5.2 \%    & 4.5 \% \\ 
\hline
\end{tabular}
\end{center}
\caption{\label{tab:zdc} 
The number of events that satisfy various selections on the number of neutrons detected 
in the ZDCs. 0n0n corresponds to no neutrons emitted in any direction; Xn to 
at least one neutron emitted in any direction; 0nXn to no neutrons in one direction and
at least one neutron in the other direction; XnXn to at least one neutron in both  
directions. For the relative yield the systematic error is estimated, as explained in the text. 
}
\end{table}

\section{Conclusions}

%
%

The first LHC measurement on coherent photoproduction of $\rho^0$ 
in Pb--Pb collisions at $\sqrt{s_{\rm NN}} = 2.76$~TeV has been presented. 
Comparisons with model calculations show that the measured cross section is in agreement with
the predictions by STARLIGHT~\cite{Klein:1999qj} and Gon\c{c}alves and Machado
(GM)~\cite{Goncalves:2011vf}, despite the idiosyncrasies in these models mentioned above. 
The Glauber-Donnachie-Landshoff (GDL) model~\cite{Frankfurt:2002wc,Rebyakova:2011vf} overpredicts the cross
section by about a factor of two. Comparisons with results from Au--Au collisions at RHIC energies indicate that
this factor of two difference is independent of collision energy in the range
$\sqrt{s_{\rm NN}} = 62.4$--$2760$~GeV. In a recent preprint, it is argued that inelastic nuclear
shadowing combined with the inclusion of intermediate states with higher mass in the $\gamma$--vector
meson transtition could explain the discrepancy~\cite{Frankfurt:2015cwa}. Regardless of whether this is
the correct explanation or not, it indicates that non-trivial corrections to the $\rho^0$ photoproduction
cross section may become important at high photon energies. 

The relative yields for different fragmentation scenarios 
are found to be in agreement with predictions from the STARLIGHT and GDL models. 
This is important not only to confirm the assumptions in the two models 
but also because some experiments, e.g. PHENIX~\cite{Afanasiev:2009hy}, have 
relied on a ZDC signal to trigger on ultra-peripheral collisions. To be able to relate such 
measurements to a photonuclear cross section, it is imperative that the probabilities for
exchange of multiple photons are well understood. 

The total cross 
section is found to be about half the total hadronic inelastic cross section. This is an 
increase of about a factor of 5 from Au--Au collisions at $\sqrt{s_{\rm NN}} = 200$~GeV, where 
the fraction was about 10\%.  If the increase of the coherent $\rho^0$ photoproduction cross 
section continues to follow STARLIGHT, one can expect it to exceed the total hadronic production cross 
section of heavy ions such as lead or gold at a $\sqrt{s_{\rm NN}}$ of about 20~TeV.

%% file: acknowledgements_jan2015.tex
The ALICE Collaboration would like to thank all its engineers and technicians for their invaluable contributions to the construction of the experiment and the CERN accelerator teams for the outstanding performance of the LHC complex.
The ALICE Collaboration gratefully acknowledges the resources and support provided by all Grid centres and the Worldwide LHC Computing Grid (WLCG) collaboration.
The ALICE Collaboration acknowledges the following funding agencies for their support in building and
running the ALICE detector:
State Committee of Science,  World Federation of Scientists (WFS)
and Swiss Fonds Kidagan, Armenia,
Conselho Nacional de Desenvolvimento Cient\'{\i}fico e Tecnol\'{o}gico (CNPq), Financiadora de Estudos e Projetos (FINEP),
Funda\c{c}\~{a}o de Amparo \`{a} Pesquisa do Estado de S\~{a}o Paulo (FAPESP);
National Natural Science Foundation of China (NSFC), the Chinese Ministry of Education (CMOE)
and the Ministry of Science and Technology of China (MSTC);
Ministry of Education and Youth of the Czech Republic;
Danish Natural Science Research Council, the Carlsberg Foundation and the Danish National Research Foundation;
The European Research Council under the European Community's Seventh Framework Programme;
Helsinki Institute of Physics and the Academy of Finland;
French CNRS-IN2P3, the `Region Pays de Loire', `Region Alsace', `Region Auvergne' and CEA, France;
German Bundesministerium fur Bildung, Wissenschaft, Forschung und Technologie (BMBF) and the Helmholtz Association;
General Secretariat for Research and Technology, Ministry of
Development, Greece;
Hungarian Orszagos Tudomanyos Kutatasi Alappgrammok (OTKA) and National Office for Research and Technology (NKTH);
Department of Atomic Energy and Department of Science and Technology of the Government of India;
Istituto Nazionale di Fisica Nucleare (INFN) and Centro Fermi -
Museo Storico della Fisica e Centro Studi e Ricerche "Enrico
Fermi", Italy;
MEXT Grant-in-Aid for Specially Promoted Research, Ja\-pan;
Joint Institute for Nuclear Research, Dubna;
National Research Foundation of Korea (NRF);
Consejo Nacional de Cienca y Tecnologia (CONACYT), Direccion General de Asuntos del Personal Academico(DGAPA), M\'{e}xico, :Amerique Latine Formation academique – European Commission(ALFA-EC) and the EPLANET Program
(European Particle Physics Latin American Network)
Stichting voor Fundamenteel Onderzoek der Materie (FOM) and the Nederlandse Organisatie voor Wetenschappelijk Onderzoek (NWO), Netherlands;
Research Council of Norway (NFR);
National Science Centre, Poland;
Ministry of National Education/Institute for Atomic Physics and Consiliul Naţional al Cercetării Ştiinţifice - Executive Agency for Higher Education Research Development and Innovation Funding (CNCS-UEFISCDI) - Romania;
Ministry of Education and Science of Russian Federation, Russian
Academy of Sciences, Russian Federal Agency of Atomic Energy,
Russian Federal Agency for Science and Innovations and The Russian
Foundation for Basic Research;
Ministry of Education of Slovakia;
Department of Science and Technology, South Africa;
Centro de Investigaciones Energeticas, Medioambientales y Tecnologicas (CIEMAT), E-Infrastructure shared between Europe and Latin America (EELA), Ministerio de Econom\'{i}a y Competitividad (MINECO) of Spain, Xunta de Galicia (Conseller\'{\i}a de Educaci\'{o}n),
Centro de Aplicaciones Tecnológicas y Desarrollo Nuclear (CEA\-DEN), Cubaenerg\'{\i}a, Cuba, and IAEA (International Atomic Energy Agency);
Swedish Research Council (VR) and Knut $\&$ Alice Wallenberg
Foundation (KAW);
Ukraine Ministry of Education and Science;
United Kingdom Science and Technology Facilities Council (STFC);
The United States Department of Energy, the United States National
Science Foundation, the State of Texas, and the State of Ohio;
Ministry of Science, Education and Sports of Croatia and  Unity through Knowledge Fund, Croatia.
Council of Scientific and Industrial Research (CSIR), New Delhi, India

%% file: Alice_Authorlist_2015-Mar-13_mod.tex


\begingroup
\small
\begin{flushleft}
J.~Adam\Irefn{org39}\And
D.~Adamov\'{a}\Irefn{org82}\And
M.M.~Aggarwal\Irefn{org86}\And
G.~Aglieri Rinella\Irefn{org36}\And
M.~Agnello\Irefn{org110}\And
N.~Agrawal\Irefn{org47}\And
Z.~Ahammed\Irefn{org130}\And
S.U.~Ahn\Irefn{org67}\And
I.~Aimo\Irefn{org93}\textsuperscript{,}\Irefn{org110}\And
S.~Aiola\Irefn{org135}
% ...
\renewcommand\labelenumi{\textsuperscript{\theenumi}~}

\section*{Affiliation notes}
\renewcommand\theenumi{\roman{enumi}}
\begin{Authlist}
\item \Adef{0}Deceased
\item \Adef{idp5852480}{Also at: University of Kansas, Lawrence, Kansas, United States}
\end{Authlist}

\section*{Collaboration Institutes}
\renewcommand\theenumi{\arabic{enumi}~}
\begin{Authlist}

\item \Idef{org1}A.I. Alikhanyan National Science Laboratory (Yerevan Physics Institute) Foundation, Yerevan, Armenia
\item \Idef{org2}Benem\'{e}rita Universidad Aut\'{o}noma de Puebla, Puebla, Mexico
\item \Idef{org3}Bogolyubov Institute for Theoretical Physics, Kiev, Ukraine
\item \Idef{org4}Bose Institute, Department of Physics and Centre for Astroparticle Physics and Space Science (CAPSS), Kolkata, India
\item \Idef{org5}Budker Institute for Nuclear Physics, Novosibirsk, Russia
\item \Idef{org6}California Polytechnic State University, San Luis Obispo, California, United States
\item \Idef{org7}Central China Normal University, Wuhan, China
\item \Idef{org8}Centre de Calcul de l'IN2P3, Villeurbanne, France
\item \Idef{org9}Centro de Aplicaciones Tecnol\'{o}gicas y Desarrollo Nuclear (CEADEN), Havana, Cuba
\item \Idef{org10}Centro de Investigaciones Energ\'{e}ticas Medioambientales y Tecnol\'{o}gicas (CIEMAT), Madrid, Spain
\item \Idef{org11}Centro de Investigaci\'{o}n y de Estudios Avanzados (CINVESTAV), Mexico City and M\'{e}rida, Mexico
\item \Idef{org12}Centro Fermi - Museo Storico della Fisica e Centro Studi e Ricerche ``Enrico Fermi'', Rome, Italy
\item \Idef{org13}Chicago State University, Chicago, Illinois, USA
\item \Idef{org14}China Institute of Atomic Energy, Beijing, China
\item \Idef{org15}Commissariat \`{a} l'Energie Atomique, IRFU, Saclay, France
\item \Idef{org16}COMSATS Institute of Information Technology (CIIT), Islamabad, Pakistan
\item \Idef{org17}Departamento de F\'{\i}sica de Part\'{\i}culas and IGFAE, Universidad de Santiago de Compostela, Santiago de Compostela, Spain
\item \Idef{org18}Department of Physics and Technology, University of Bergen, Bergen, Norway
\item \Idef{org19}Department of Physics, Aligarh Muslim University, Aligarh, India
\item \Idef{org20}Department of Physics, Ohio State University, Columbus, Ohio, United States
\item \Idef{org21}Department of Physics, Sejong University, Seoul, South Korea
\item \Idef{org22}Department of Physics, University of Oslo, Oslo, Norway
\item \Idef{org23}Dipartimento di Elettrotecnica ed Elettronica del Politecnico, Bari, Italy
\item \Idef{org24}Dipartimento di Fisica dell'Universit\`{a} 'La Sapienza' and Sezione INFN Rome, Italy
\item \Idef{org25}Dipartimento di Fisica dell'Universit\`{a} and Sezione INFN, Cagliari, Italy
\item \Idef{org26}Dipartimento di Fisica dell'Universit\`{a} and Sezione INFN, Trieste, Italy
\item \Idef{org27}Dipartimento di Fisica dell'Universit\`{a} and Sezione INFN, Turin, Italy
\item \Idef{org28}Dipartimento di Fisica e Astronomia dell'Universit\`{a} and Sezione INFN, Bologna, Italy
\item \Idef{org29}Dipartimento di Fisica e Astronomia dell'Universit\`{a} and Sezione INFN, Catania, Italy
\item \Idef{org30}Dipartimento di Fisica e Astronomia dell'Universit\`{a} and Sezione INFN, Padova, Italy
\item \Idef{org31}Dipartimento di Fisica `E.R.~Caianiello' dell'Universit\`{a} and Gruppo Collegato INFN, Salerno, Italy
\item \Idef{org32}Dipartimento di Scienze e Innovazione Tecnologica dell'Universit\`{a} del  Piemonte Orientale and Gruppo Collegato INFN, Alessandria, Italy
\item \Idef{org33}Dipartimento Interateneo di Fisica `M.~Merlin' and Sezione INFN, Bari, Italy
\item \Idef{org34}Division of Experimental High Energy Physics, University of Lund, Lund, Sweden
\item \Idef{org35}Eberhard Karls Universit\"{a}t T\"{u}bingen, T\"{u}bingen, Germany
\item \Idef{org36}European Organization for Nuclear Research (CERN), Geneva, Switzerland
\item \Idef{org37}Faculty of Engineering, Bergen University College, Bergen, Norway
\item \Idef{org38}Faculty of Mathematics, Physics and Informatics, Comenius University, Bratislava, Slovakia
\item \Idef{org39}Faculty of Nuclear Sciences and Physical Engineering, Czech Technical University in Prague, Prague, Czech Republic
\item \Idef{org40}Faculty of Science, P.J.~\v{S}af\'{a}rik University, Ko\v{s}ice, Slovakia
\item \Idef{org41}Faculty of Technology, Buskerud and Vestfold University College, Vestfold, Norway
\item \Idef{org42}Frankfurt Institute for Advanced Studies, Johann Wolfgang Goethe-Universit\"{a}t Frankfurt, Frankfurt, Germany
\item \Idef{org43}Gangneung-Wonju National University, Gangneung, South Korea
\item \Idef{org44}Gauhati University, Department of Physics, Guwahati, India
\item \Idef{org45}Helsinki Institute of Physics (HIP), Helsinki, Finland
\item \Idef{org46}Hiroshima University, Hiroshima, Japan
\item \Idef{org47}Indian Institute of Technology Bombay (IIT), Mumbai, India
\item \Idef{org48}Indian Institute of Technology Indore, Indore (IITI), India
\item \Idef{org49}Inha University, Incheon, South Korea
\item \Idef{org50}Institut de Physique Nucl\'eaire d'Orsay (IPNO), Universit\'e Paris-Sud, CNRS-IN2P3, Orsay, France
\item \Idef{org51}Institut f\"{u}r Informatik, Johann Wolfgang Goethe-Universit\"{a}t Frankfurt, Frankfurt, Germany
\item \Idef{org52}Institut f\"{u}r Kernphysik, Johann Wolfgang Goethe-Universit\"{a}t Frankfurt, Frankfurt, Germany
\item \Idef{org53}Institut f\"{u}r Kernphysik, Westf\"{a}lische Wilhelms-Universit\"{a}t M\"{u}nster, M\"{u}nster, Germany
\item \Idef{org54}Institut Pluridisciplinaire Hubert Curien (IPHC), Universit\'{e} de Strasbourg, CNRS-IN2P3, Strasbourg, France
\item \Idef{org55}Institute for Nuclear Research, Academy of Sciences, Moscow, Russia
\item \Idef{org56}Institute for Subatomic Physics of Utrecht University, Utrecht, Netherlands
\item \Idef{org57}Institute for Theoretical and Experimental Physics, Moscow, Russia
\item \Idef{org58}Institute of Experimental Physics, Slovak Academy of Sciences, Ko\v{s}ice, Slovakia
\item \Idef{org59}Institute of Physics, Academy of Sciences of the Czech Republic, Prague, Czech Republic
\item \Idef{org60}Institute of Physics, Bhubaneswar, India
\item \Idef{org61}Institute of Space Science (ISS), Bucharest, Romania
\item \Idef{org62}Instituto de Ciencias Nucleares, Universidad Nacional Aut\'{o}noma de M\'{e}xico, Mexico City, Mexico
\item \Idef{org63}Instituto de F\'{\i}sica, Universidad Nacional Aut\'{o}noma de M\'{e}xico, Mexico City, Mexico
\item \Idef{org64}iThemba LABS, National Research Foundation, Somerset West, South Africa
\item \Idef{org65}Joint Institute for Nuclear Research (JINR), Dubna, Russia
\item \Idef{org66}Konkuk University, Seoul, South Korea
\item \Idef{org67}Korea Institute of Science and Technology Information, Daejeon, South Korea
\item \Idef{org68}KTO Karatay University, Konya, Turkey
\item \Idef{org69}Laboratoire de Physique Corpusculaire (LPC), Clermont Universit\'{e}, Universit\'{e} Blaise Pascal, CNRS--IN2P3, Clermont-Ferrand, France
\item \Idef{org70}Laboratoire de Physique Subatomique et de Cosmologie, Universit\'{e} Grenoble-Alpes, CNRS-IN2P3, Grenoble, France
\item \Idef{org71}Laboratori Nazionali di Frascati, INFN, Frascati, Italy
\item \Idef{org72}Laboratori Nazionali di Legnaro, INFN, Legnaro, Italy
\item \Idef{org73}Lawrence Berkeley National Laboratory, Berkeley, California, United States
\item \Idef{org74}Lawrence Livermore National Laboratory, Livermore, California, United States
\item \Idef{org75}Moscow Engineering Physics Institute, Moscow, Russia
\item \Idef{org76}National Centre for Nuclear Studies, Warsaw, Poland
\item \Idef{org77}National Institute for Physics and Nuclear Engineering, Bucharest, Romania
\item \Idef{org78}National Institute of Science Education and Research, Bhubaneswar, India
\item \Idef{org79}Niels Bohr Institute, University of Copenhagen, Copenhagen, Denmark
\item \Idef{org80}Nikhef, National Institute for Subatomic Physics, Amsterdam, Netherlands
\item \Idef{org81}Nuclear Physics Group, STFC Daresbury Laboratory, Daresbury, United Kingdom
\item \Idef{org82}Nuclear Physics Institute, Academy of Sciences of the Czech Republic, \v{R}e\v{z} u Prahy, Czech Republic
\item \Idef{org83}Oak Ridge National Laboratory, Oak Ridge, Tennessee, United States
\item \Idef{org84}Petersburg Nuclear Physics Institute, Gatchina, Russia
\item \Idef{org85}Physics Department, Creighton University, Omaha, Nebraska, United States
\item \Idef{org86}Physics Department, Panjab University, Chandigarh, India
\item \Idef{org87}Physics Department, University of Athens, Athens, Greece
\item \Idef{org88}Physics Department, University of Cape Town, Cape Town, South Africa
\item \Idef{org89}Physics Department, University of Jammu, Jammu, India
\item \Idef{org90}Physics Department, University of Rajasthan, Jaipur, India
\item \Idef{org91}Physik Department, Technische Universit\"{a}t M\"{u}nchen, Munich, Germany
\item \Idef{org92}Physikalisches Institut, Ruprecht-Karls-Universit\"{a}t Heidelberg, Heidelberg, Germany
\item \Idef{org93}Politecnico di Torino, Turin, Italy
\item \Idef{org94}Purdue University, West Lafayette, Indiana, United States
\item \Idef{org95}Pusan National University, Pusan, South Korea
\item \Idef{org96}Research Division and ExtreMe Matter Institute EMMI, GSI Helmholtzzentrum f\"ur Schwerionenforschung, Darmstadt, Germany
\item \Idef{org97}Rudjer Bo\v{s}kovi\'{c} Institute, Zagreb, Croatia
\item \Idef{org98}Russian Federal Nuclear Center (VNIIEF), Sarov, Russia
\item \Idef{org99}Russian Research Centre Kurchatov Institute, Moscow, Russia
\item \Idef{org100}Saha Institute of Nuclear Physics, Kolkata, India
\item \Idef{org101}School of Physics and Astronomy, University of Birmingham, Birmingham, United Kingdom
\item \Idef{org102}Secci\'{o}n F\'{\i}sica, Departamento de Ciencias, Pontificia Universidad Cat\'{o}lica del Per\'{u}, Lima, Peru
\item \Idef{org103}Sezione INFN, Bari, Italy
\item \Idef{org104}Sezione INFN, Bologna, Italy
\item \Idef{org105}Sezione INFN, Cagliari, Italy
\item \Idef{org106}Sezione INFN, Catania, Italy
\item \Idef{org107}Sezione INFN, Padova, Italy
\item \Idef{org108}Sezione INFN, Rome, Italy
\item \Idef{org109}Sezione INFN, Trieste, Italy
\item \Idef{org110}Sezione INFN, Turin, Italy
\item \Idef{org111}SSC IHEP of NRC Kurchatov institute, Protvino, Russia
\item \Idef{org112}SUBATECH, Ecole des Mines de Nantes, Universit\'{e} de Nantes, CNRS-IN2P3, Nantes, France
\item \Idef{org113}Suranaree University of Technology, Nakhon Ratchasima, Thailand
\item \Idef{org114}Technical University of Split FESB, Split, Croatia
\item \Idef{org115}The Henryk Niewodniczanski Institute of Nuclear Physics, Polish Academy of Sciences, Cracow, Poland
\item \Idef{org116}The University of Texas at Austin, Physics Department, Austin, Texas, USA
\item \Idef{org117}Universidad Aut\'{o}noma de Sinaloa, Culiac\'{a}n, Mexico
\item \Idef{org118}Universidade de S\~{a}o Paulo (USP), S\~{a}o Paulo, Brazil
\item \Idef{org119}Universidade Estadual de Campinas (UNICAMP), Campinas, Brazil
\item \Idef{org120}University of Houston, Houston, Texas, United States
\item \Idef{org121}University of Jyv\"{a}skyl\"{a}, Jyv\"{a}skyl\"{a}, Finland
\item \Idef{org122}University of Liverpool, Liverpool, United Kingdom
\item \Idef{org123}University of Tennessee, Knoxville, Tennessee, United States
\item \Idef{org124}University of the Witwatersrand, Johannesburg, South Africa
\item \Idef{org125}University of Tokyo, Tokyo, Japan
\item \Idef{org126}University of Tsukuba, Tsukuba, Japan
\item \Idef{org127}University of Zagreb, Zagreb, Croatia
\item \Idef{org128}Universit\'{e} de Lyon, Universit\'{e} Lyon 1, CNRS/IN2P3, IPN-Lyon, Villeurbanne, France
\item \Idef{org129}V.~Fock Institute for Physics, St. Petersburg State University, St. Petersburg, Russia
\item \Idef{org130}Variable Energy Cyclotron Centre, Kolkata, India
\item \Idef{org131}Vin\v{c}a Institute of Nuclear Sciences, Belgrade, Serbia
\item \Idef{org132}Warsaw University of Technology, Warsaw, Poland
\item \Idef{org133}Wayne State University, Detroit, Michigan, United States
\item \Idef{org134}Wigner Research Centre for Physics, Hungarian Academy of Sciences, Budapest, Hungary
\item \Idef{org135}Yale University, New Haven, Connecticut, United States
\item \Idef{org136}Yonsei University, Seoul, South Korea
\item \Idef{org137}Zentrum f\"{u}r Technologietransfer und Telekommunikation (ZTT), Fachhochschule Worms, Worms, Germany
\end{Authlist}
\endgroup